\documentstyle[epsfig]{aipproc}

\def\etal{{et al. }}
\def\HH{${\rm {H_2}}\,\,$}

\def\cm{{\rm cm}}

\def\gs{\mathrel{\raise1.16pt\hbox{$>$}\kern-7.0pt
\lower3.06pt\hbox{{$\scriptstyle \sim$}}}}
\def\ls{\mathrel{\raise1.16pt\hbox{$<$}\kern-7.0pt
\lower3.06pt\hbox{{$\scriptstyle \sim$}}}}
\def\gtsima{$\; \buildrel > \over \sim \;$}
\def\ltsima{$\; \buildrel < \over \sim \;$}
\def\prosima{$\; \buildrel \propto \over \sim \;$}
\def\gsim{\lower.5ex\hbox{\gtsima}}
\def\lsim{\lower.5ex\hbox{\ltsima}}
\def\simgt{\lower.5ex\hbox{\gtsima}}
\def\simlt{\lower.5ex\hbox{\ltsima}}
\def\simpr{\lower.5ex\hbox{\prosima}}

\def\pp{\noindent\parshape 2 0truecm 17truecm 2truecm 15truecm}
\def\rf#1;#2;#3;#4 {\par\pp#1, #2, #3, #4. \par}

\def\pr{\ref@jnl{Phys.Rev}}

\def\href#1;#2 {{\bf #1} : {\em #2}}

\def\ssubsection#1 {\subsection{\sl #1}}

\def\beq#1{\begin{equation}\label{#1}}
\def\eeq{\end{equation}}
\def\beqa#1{\begin{eqnarray}\label{#1}}
\def\eeqa{\end{eqnarray}}
\def\eq#1{equation~(\ref{#1})}

\def\tento#1{\times 10^{#1}}

\def\Ms{\ M_{\odot}}

\def\K{{\rm \ K}}
\def\s{{\rm \ s}}

\def\ergs{{\rm \ erg}}

\def\cm{{\rm \ cm}}

\def\kpc{{\rm \ kpc}}

\def\HH{H$_2$ }
\def\H2p{H$_2^+$ }

\def\Hm{H$^-$ }

\def\mH2p{H_2^+}

\begin{document}

\title{\Large\sc Primordial Star Forming Regions in a CDM Universe}

\author{Yu Zhang,  Michael L. Norman, Peter Anninos, Tom Abel }
\vspace{1cm}
\address{\it  Laboratory for Computational Astrophysics  \\
University of Illinois at Urbana--Champaign \\
405 N. Mathews Ave., Urbana, IL 61801}

\maketitle


\begin{abstract}
    
  We developed a three--dimensional 2--level hierarchical cosmological
  code  with  a   realistic and  robust   treatment of  multi--species
  non--equilibrium ionization and radiative cooling processes, and use
  it to investigate  primordial star  forming regions  that  originate
  from high--$\sigma$   perturbations in  a  standard  CDM   dominated
  cosmology.  We find it  is possible to produce gravitationally bound
  and cooled structures   at very high   redshift ($z  \sim 40$)  with
  baryonic masses as small as  $\sim 10^3\Ms$.  The molecular hydrogen
  formation in these   small scale structures  follows  very  well the
  analytical predictions of Abel (1995) and  Tegmark \etal (1996).  We
  also discuss the minimum mass that cosmological structures must have
  in order to be able to cool and collapse.

\end{abstract}


\section{Introduction}
\label{sec:introduction}

Models for  structure   formation are  based  on the  growth of  small
primordial  density   fluctuations by gravitational   instability on a
homogeneously expanding  background universe.  Depending on the nature
of    the dark matter   and  whether the  primordial fluctuations were
adiabatic or isothermal, the first mass  scale to collapse could be as
small as one solar mass (very heavy cold dark  matter particles) or as
high  as $\sim 10^{12}M_\odot$  (hot  dark matter scenarios).  In  CDM
cosmogonies the  fluctuation spectrum at small  wavelengths has only a
logarithmic  dependence    for  mass scales      smaller than    $\sim
10^8M_\odot$, which indicates  that the  small scale fluctuations   in
this model collapse nearly simultaneously in time.  This leads to very
complex dynamics during the formation of these structures, that can be
studied accurately only  by using realistic  numerical computations to
model the  fluid motion and micro--physical processes  as  well as the
dark matter component.

We  have recently been able to  develop methods that allow
us   to study  the problem in    three  dimensions \cite{AZAN96}.   We
describe this code  briefly in section \ref{3D}  but  first review the
process of \HH formation during  small scale structure collapse in the
following section.  A more   extensive  discussion of our  results  is
given in a separate publication \cite{AANZ96}.

\section{Molecular Chemistry and Cooling}\label{sec:H2formation}

The  cooling in   small scale  fluctuations  is dominated  by the
rotational/vibrational modes of hydrogen molecules.  
In  primordial gas  at  low  temperatures   ($\lsim 6000$K)
molecular hydrogen can not  be  destroyed efficiently unless there is a
radiation flux higher  than  $\sim 3  \times 10^{-26}   \ergs \cm^{-2}
\s^{-1} $ in the Lyman Werner Bands.  Once self-shielding is important,
even higher fluxes  would be needed.   The dominating H$_2$  producing
gas phase reaction is the dissociative attachment reaction:
$\rm
\ \ H^- \  + \  H \  \rightarrow \  H_2 \  + \  e^-.
$ In the absence of an external UV background one can integrate the rate
equations  to find the molecular  hydrogen fraction  formed during the
collapse  of  primordial  gas clouds    with neutral  hydrogen  number
density $n_H$, temperature $T$, and  initial free electron fraction
$x_0$ to be \cite{A95,TSRBAP96}:
\begin{eqnarray}\label{equ:fH2}
f_{H_2}(t) - f_{H_2}(0) &=&  \frac{k_{PA}}{k_{r}} \ln(x_0
n_H k_{r} t + 1) =  10^{-8}T^{1.53} \ln(t/t_{r}^0 + 1),
\end{eqnarray}
where $k_{PA}$,  and $k_{r}$ denote  the rates for photo-attachment to
\Hm and recombination of  hydrogen,  respectively.  The production  of
H$_2$ only depends logarithmically  on time with  a typical time scale
of one initial recombination  time.  The temperature dependence is due
to the ratio of recombination and \Hm formation time scales, which is a
measure of  the  number of electrons   available to  produce   \Hm.  A
typical \HH fraction of $\sim 10^{-3}$ is produced during the collapse
of structures with  virial temperatures greater  than $ 10^3 \K$.  For
initial (virial) temperatures higher than $6000\K$ the charge exchange
with protons will efficiently destroy H$_2$, and \eq{equ:fH2} will not
be applicable.  However, during the collapse  of clouds with such high
virial  temperatures  the final \HH   fraction  will, nevertheless, be
$f_{H_2}(T\sim 6000\K ) \sim few \tento{-3}$ \cite{AAZN96}.

\section{Numerical Results and Discussion}
\label{3D}

We   achieve high spatial   and  mass resolution  with the  two--level
hierarchical three--dimensional code (HERCULES) that we have developed
for cosmology \cite{ANC94,AZAN96}.  This code  is designed to simulate
structure   formation in an expanding   dark matter dominated universe
with Newtonian gravity, multi--fluid hydrodynamics, radiative cooling,
non--equilibrium chemistry and external radiation fields.
Furthermore,  the  code   independently  evolves  the following   nine
species:  neutral      hydrogen   $H$,  ionized      hydrogen   $H^+$,
negatively--charged hydrogen $H^-$,  hydrogen molecules $H_2$, ionized
hydrogen  molecules ${H_2}^+$,   neutral helium $He$,  singly--ionized
helium  $He^+$,  doubly--ionized helium  $He^{++}$ and  free electrons
$e^-$.  The   28  most important chemical   rate  equations (including
radiation processes) are solved in non--equilibrium for the abundances
of   each of the  nine species.   The reaction rates  and an extensive
discussion of the chemistry  model are provided  in \cite{AAZN96}.  We
have also implemented a comprehensive  model for the radiative cooling
of the gas  that  includes   atomic line excitation,    recombination,
collisional  ionization,   free--free  transitions,    molecular  line
excitations, and Compton scattering of the cosmic background radiation
(CBR) by electrons.

We apply our code to  high redshift pre--galactic structure  formation
and evolution, investigating  specifically  the collapse of  the first
high--$\sigma$ bound objects with total   masses in the range $10^5  -
10^9 M_\odot$.     Our   model  background   spacetime    is  a   flat
($\Omega_0=1$)    cold dark matter    dominated  universe  with Hubble
constant     $H_0=50$     km~s$^{-1}$~Mpc$^{-1}$,   baryonic  fraction
$\Omega_B=0.06$, and a hydrogen mass fraction of  76\% .  The baryonic
matter is composed of hydrogen  and helium in  cosmic abundance with a
hydrogen   mass    fraction of  76\%   and   ratio  of  specific heats
$\gamma=5/3$.   The   initial data for the   baryonic  and dark matter
perturbations  is the  Harrison--Zel'dovich  power spectrum  modulated
with  a CDM transfer  function  and normalized   to the cluster  scale
$\sigma_{8h^{-1}}=0.7$.  The data is  initialized at  redshift $z=100$
using Bertschinger's \cite{B94} constrained realization procedure to
construct 3  and  4$\sigma$ fluctuations  in cubes  of comoving length
1024kpc, 512kpc, and 128kpc,  with total masses of $7.5\tento{10}\Ms$,
$9.3\tento{9}\Ms$, and $1.5\tento{8}\Ms$, respectively.

We reproduced  the work  of Tegmark  \etal 1996  with the same cooling
function \cite{LS83} we  have  used in our  cosmological hydrodynamics
code  so that we   can   compare their findings  directly  to   our 3D
numerical results.   (We note that Tegmark  \etal used a modified form
of the Hollenbach and McKee (1979) \HH cooling function.)
\begin{figure}[ht]
\centerline{\epsfig{file=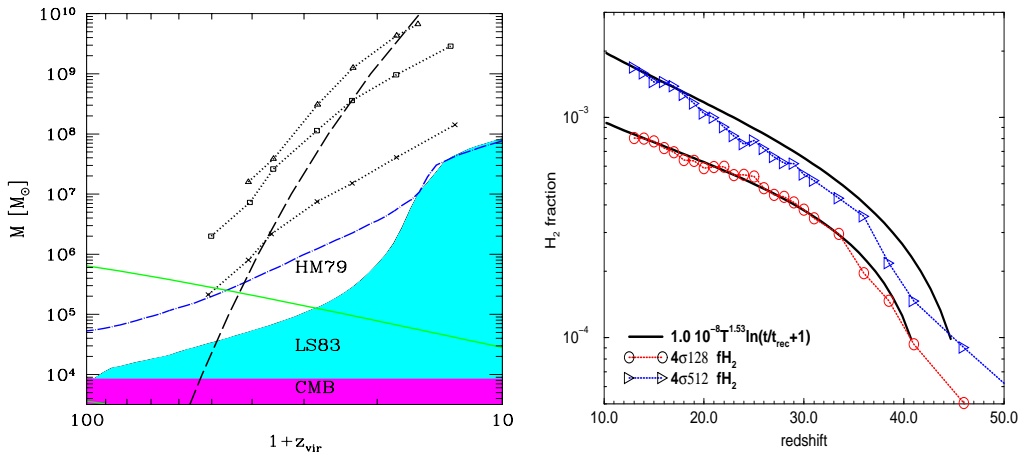,height=7cm,width=14cm}}
\caption{ \small{   (a)  Collapsed mass vs.  redshift.   The  dark
    shaded  region depicts the    mass  scale for  which  the   virial
    temperature  equals the CMB   temperature.  Only  above the  light
    shaded area labeled  with LS83  are  structures able  to collapse. 
    The  dotted lines show $M_{200}$  from our $4\sigma$ 128, 512, and
    1024kpc  (crosses, squares,  triangles)  runs.  The  dashed upward
    sloping line  is the  BBKS CDM  spectrum scaled  appropriately for
    $4\sigma$  peaks.  The gray solid  downward sloping  line  is the Jeans
    Mass at  $18 \pi^2$ the background  density.  The dot--dashed line
    shows the original   delimiting  line computed by  Tegmark   \etal
    (1996) which was  based on a  modified form of the Hollenbach  and
    McKee (1979)  \HH  cooling function.  (b) The   molecular hydrogen
    fraction  vs.  redshift in the  densest  zone of the  $4\sigma128$
    (open circles) and the  $4\sigma512$ simulations (triangles).  The
    thick solid lines is the solution (\ref{equ:fH2}) initialized with
    the appropriate free electron densities, initial \HH fraction, and
    temperature. } } \label{Teg96}
\end{figure}
Figure 1a is analogous to Fig.   6 of Tegmark  \etal 1996 but includes
the mass evolution in     our  numerical results for   the   4$\sigma$
perturbations.  The dotted lines are found by adding up the total mass
$M_{200}$ found in cells with dark matter overdensities exceeding 200.
It is obvious that the use of a different cooling  function has a very
strong  influence on the predicted mass  that  can collapse.  Although,
the quantitative results are very  different, the shape for these  two
different  $M_c(z)$  curves is rather similar   at redshifts $30<  z <
100$.  The  slopes are consistent  with $M_c  \propto  (1+z)^{-3/2}$
indicating  a    constant virial  temperature  since  $T_{vir} \propto
M^{2/3} (1+z)$.  A   constant  virial temperature in  turn   implies a
roughly  constant final \HH fraction   given by \eq{equ:fH2}.  For the
case in Tegmark \etal, the virial temperature in  that regime is $\sim
1000\K$ which yields $f_{H_2} \sim 4 \times 10^{-4}$ which they argued
to  be roughly a constant  universal value which,  if exceeded, allows
the cloud to collapse at its free fall rate.  Using the Lepp and Shull
\HH cooling function   we find that the  virial  temperature needed to
fulfill the Tegmark \etal requirement  for collapse is $\approx 200\K$
for redshifts $>30$.  This translates  to a molecular fraction of only
$\sim 3 \times 10^{-5}$.  Our simulations, however, show that, although
we used  the Lepp and Shull  cooling function, the
\HH fraction at the  time when the baryons  are collapsing into the DM
potential wells is also about $\sim 5 \times 10^{-4}$.  Hence, we find
roughly the same   critical \HH fraction as  derived  by Tegmark \etal
(1996) even with a very different cooling function.


In  Figure  1b  we    test  \eq{equ:fH2} against   results   from  the
$4\sigma128\kpc$ (open  circles) and the $4\sigma512\kpc$ simulations. 
The fit is  astonishingly  good, although the initial  temperature (or
redshift) is  somewhat  difficult to pick  out in  this case since the
heating due to adiabatic compression  is slow and the initial (virial)
temperature is not as   well defined as  in  the case of  collapse  on
larger mass scales.  Analyzing the time derivative of \eq{equ:fH2} one
finds for large  times that $\dot{f_{H_2}}  \propto T^{1.53}/t$ which,
for the spherical collapse model, translates to $\dot{f_{H_2}} \propto
H_0 M (1+z_{vir})^{1.53}$ when   we compare the slopes   for different
mass scales at the present time.  This explains that the divergence of
the two graphs in Figure 1b for low redshifts is due to differences in
collapse mass and redshift.

It has    been stressed by   various authors  that  early  small scale
structure might     influence the    entire  pregalactic medium    and
subsequently play an important role  for structure formation on larger
mass scales (e.g.   \cite{CR86}).  They are   in principle capable  of
ionizing a  large  fraction of  the pregalactic medium   as well as to
enrich  it  with metals.  We  hope to  achieve  the required dynamical
range  in future   work  to  estimate  the   IMF using  adaptive  mesh
refinement  techniques  and so be  able  to  quantify  the feedback of
collapsing small scale structure.  \ 

We happily acknowledge  discussions with M.J.   Rees and Max Tegmark.  
This work is done under the  auspices of the Grand Challenge Cosmology
Consortium (GC3) and supported in part by  NSF grant ASC-9318185.  The
simulations   were  performed on  the  CRAY-C90 at  the   PSC, and the
CONVEX-3880 at the NCSA.

\end{document}